\documentclass[ceqn, prd, showpacs, showkeys, prd, onecolumn, nofootinbib, amsmath, amsfonts, amssymb]{revtex4-1}

\usepackage{amssymb,amsmath,mathtools,xcolor,graphicx,xspace,colortbl,ragged2e,rotating} %
\usepackage{textcomp}
\usepackage{amsfonts}  
\usepackage{amsmath}  
\usepackage{amssymb}  
\usepackage{boxedminipage}  
\usepackage{graphics}  
\usepackage{graphicx}  
\usepackage{ragged2e}  
\usepackage{tabulary}  
\usepackage{wrapfig}  
\usepackage{xcolor}  
\usepackage{varioref}  
\graphicspath{{FRACTIONAL7_1_graphics/}{FRACTIONAL7_1_tcache/}{FRACTIONAL7_1_gcache/}}
\DeclareGraphicsExtensions{.pdf,.eps,.ps,.png,.jpg,.jpeg}

\begin{document}
\title{NEWTONIAN FRACTIONAL-DIMENSION GRAVITY AND GALAXIES WITHOUT DARK MATTER
}
\author{Gabriele U. Varieschi
}

\email[E-mail me at: ]{Gabriele.Varieschi@lmu.edu
}
\homepage[Visit: ]{https://gvarieschi.lmu.build
}
\affiliation{Department of Physics, Loyola Marymount University, Los Angeles, CA 90045, USA
}

\date{
\today
}
\begin{abstract}
We apply Newtonian Fractional-Dimension Gravity (NFDG), an alternative gravitational model, to some notable cases of galaxies with little or no dark matter. In the case of the ultra-diffuse galaxy AGC 114905, we show that NFDG\ methods can effectively reproduce the observed rotation curve, by using a variable fractional dimension $D\left (R\right )$ as was done for other galaxies in previous studies. For AGC 114905, we obtain a variable dimension in the range $D \approx 2.2 -3.2$, but our fixed $D =3$ curve can still fit all the experimental data within their error bars. This confirms other studies indicating that the dynamics of this galaxy  can be described almost entirely by the baryonic mass distribution alone.

In the case of NGC 1052-DF2, we use an argument based on the NFDG extension of the virial theorem applied to the velocity dispersion of globular clusters showing that, in general, discrepancies between observed and predicted velocity dispersions can be attributed to an overall fractal dimension $D <3$ of the astrophysical structure considered and not to the presence of dark matter. For NGC 1052-DF2 we estimate $D \approx 2.9$, thus confirming that this galaxy almost follows standard Newtonian behavior.

We also consider the case of the Bullet Cluster merger (1E0657-56), assumed to be one of the strongest proofs of dark matter existence. A simplified but effective NFDG model of the collision shows that the observed infall velocity of this merger can be explained by a fractional dimension of the system in the range $D \simeq 2.4 -2.5$, again without using any dark matter.

\end{abstract}
\maketitle

\section{\label{sect:intro}
Introduction
}
Newtonian Fractional-Dimension Gravity (NFDG) is an alternative gravitational model introduced in 2020-2021 through a series of papers (\cite{Varieschi:2020ioh,Varieschi:2020dnd,Varieschi:2020hvp}, hereafter papers I-III, respectively), with the goal of modeling galactic dynamics without using any Dark Matter (DM) component. The main assumption of NFDG is to consider galaxies, and possibly other astrophysical objects, as fractal structures characterized by a fractional-dimension function $D\left (R\right )$, where $D$ represents a variable Hausdorff-type fractal dimension, which typically depends on the radial coordinate of the structure being studied and can assume any positive real value (usually, $1 <D \lesssim 3$).\protect\footnote{
For a general overview of Newtonian Fractional-Dimension Gravity, see also the NFDG website \cite{Varieschi:webpage}.
} 

An intuitive understanding of the concept of fractal dimension is based on fractal geometry, as originally introduced by B. Mandelbrot \cite{MAN83}, where a fractal object containing a detailed structure at small scales is characterized by a fractal dimension that can differ from the standard topological dimension (i.e., different from the standard integer dimension $D=1,2,3$). Fractal phenomena, displaying self-similarity and fractal dimensions, are being uncovered in stellar systems, galaxies, and other astrophysical structures \cite{Baryshev:2002tn}, thus leading to applications of fractals to astrophysics and cosmology.

The main motivation for considering a theory of gravity based on a fractional dimension $1 <D \lesssim 3$ is that it leads to a gravitational potential of the form $\sim 1/r^{D-2}$ (or $\sim \ln r$ for $D=2$), as in Eq. (\ref{eq2.1}) below. This gravitational potential interpolates naturally between the standard Newtonian one ($\sim 1/r$, for $D=3$) and those yielding flat galactic rotation curves, without the need to introduce any DM component. We also note that NFDG should be considered \cite{Varieschi:2020ioh,Varieschi:2020dnd,Varieschi:2020hvp} as a pure modification of the laws of gravity, as briefly outlined in Sect. \ref{sect:agc}. Therefore, NFDG will not affect other fundamental interactions such as electromagnetism, or the way electromagnetic waves propagate in the Universe, etc.

In papers I-III, this model was shown to work effectively for general structures characterized by spherical or axial symmetries, as well as for three notable cases of rotationally-supported galaxies (NGC 6503, NGC 7814, and NGC 3741), by fitting their rotation curves without using any DM\  contribution. In a subsequent paper IV \cite{Varieschi:2021rzk}, a relativistic version of the model was presented, while in our paper V \cite{Varieschi:2022mid} four more galaxies were analyzed (NGC 5033, NGC 6674, NGC 5055, NGC 1090), in relation to the so-called External Field Effect (EFE), and accurate fits to their rotation velocity data were obtained also in these cases.

In all these previous papers, we discussed possible points of contact with other existing alternative gravities (Modified Newtonian Dynamics - MOND \cite{Milgrom:1983ca,Milgrom:1983pn,Milgrom:1983zz,Bekenstein:1984tv,Milgrom:2001ny,2002ARA&A..40..263S,Famaey:2011kh,Skordis:2020eui,Merritt:2020pwe}, Conformal Gravity - CG \cite{Mannheim:1988dj,Mannheim:1992tr,Mannheim:2005bfa,Mannheim:2010ti,Mannheim:2010xw,Mannheim:2011ds}, Modified Gravity - MOG \cite{Moffat:2005si,Moffat:2010gt,Green:2019cqm}, etc.), as well as links with fractional gravity theories in multi-scale spacetimes \cite{Calcagni:2011sz,Calcagni:2009kc,Calcagni:2010bj,Calcagni:2013yqa,Calcagni:2020ads,Calcagni:2021ipd,Calcagni:2021aap,Calcagni:2021mmj}. In addition to these alternative theories of gravity, many other ideas have been recently explored such as gravitational confinement \cite{Barker:2023xsx}, graviton-graviton interactions \cite{Deur:2009ya}, unexpected relativistic corrections \cite{Deur:2020wlg}, including gravitomagnetic effects \cite{Cooperstock:2006dt,Ludwig:2021kea}, just to name a few.

In this short paper, we will apply our NFDG methods to the case of galaxies which appear to have little or no dark matter. In Sect. \ref{sect:agc}, we will consider the case of the ultra-diffuse galaxy (UDG) AGC 114905 \cite{PinaMancera:2021wpc,Kong:2022oyk}, while in Sect. \ref{sect:ngc} we will study another UDG, NGC 1052-DF2 \cite{vanDokkum:2018vup,vanDokkum:2018ccc,vanDokkum:2022zdd}, which is another remarkable case of a galaxy lacking dark matter. In section \ref{sect:bc}, we will also briefly analyze the Bullet Cluster merger (1E0657-56), which is usually referred to as an empirical proof of the existence of dark matter \cite{Clowe:2006eq,Lage:2014yxa}. 

Our NFDG model will prove to be effective in all these cases: for the first two, it will confirm that these galaxies can be considered to be virtually free of DM, while in the case of the Bullet Cluster we will show that its dynamics can also be explained by NFDG without including DM contributions.

\section{\label{sect:agc} NFDG and the ultra-diffuse galaxy AGC 114905
}
NFDG is based on a heuristic extension of Gauss's law for gravitation to a lower-dimensional space-time $D +1$, where $D \leq 3$  can become a non-integer space dimension. In our previous papers, we introduced a fractional-dimension Poisson equation, $\nabla _{D}^{2}\Phi  =\frac{4\pi G}{l_{0}}\rho$, where $ \nabla _{D}^{2}$ is a generalized $D$-dimensional Laplacian (see Ref. \cite{Varieschi:2020ioh}, or the more detailed discussion in Ref. \cite{Varieschi:2021rzk}), and $\rho $ represents here the rescaled mass density. The solution to this Poisson equation, for a point-like mass $m$ placed at the origin, yields the following NFDG gravitational potential \cite{Varieschi:2022mid}:
\begin{gather}\Phi _{NFDG}(r) = -\frac{2\pi ^{1 -D/2}\Gamma (D/2)\ Gm}{\left (D -2\right )l_{0}r^{D -2}}\ ;\ D \neq 2 \label{eq2.1} \\
\Phi _{NFDG}\left (r\right ) =\frac{2\ Gm}{l_{0}}\ln r\ ;\ D =2 , \nonumber \end{gather}
where $G$ is the gravitational constant, the radial coordinate $r$ is considered to be a rescaled, dimensionless coordinate, i.e., $r \rightarrow r/l_{0}$, and $l_{0}$ is an appropriate scale length, which is required for dimensional correctness in fractional gravity models.

This potential is then generalized to extended source mass distributions, in particular spherically-symmetric and axially-symmetric distributions, and used to model the three main components of galactic baryonic matter: the spherical bulge mass distribution (if present) and the cylindrical stellar disk and gas distributions. It is assumed that the space dimension $D$ can be a function of the field point coordinates, but we usually neglect its space derivatives in the computation of the NFDG gravitational field:\begin{equation}\mathbf{g}_{NFDG}\left (R\right ) = -\frac{1}{l_{0}}\frac{d\Phi _{NFDG}\left (R\right )}{dR}\widehat{\mathbf{R}} . \label{eq2.2}
\end{equation}

This field is usually computed just in the galactic disk plane and it is then expressed as a function of the (dimensionless)\ radial coordinate $R$ in the same plane, with the scale length $l_{0}$ included for dimensional correctness. Similarly, the variable dimension function $D =D\left (R\right )$ is considered a function of the same radial variable and will characterize each particular galaxy studied with NFDG methods.

Finally, the circular velocity, for stars rotating in the main galactic plane, is simply obtained from Eq. (\ref{eq2.2}) as:
\begin{equation}v_{circ}\left (R\right ) =\sqrt{l_{0}R\left \vert \mathbf{g}_{NFDG}\left (R\right )\right \vert } , \label{eq2.3}
\end{equation}where the NFDG field is computed by using the variable dimension function $D =D\left (R\right )$ mentioned above. The observed rotation velocity data for each galaxy are then fitted by plotting these circular velocities as a function of the radial distance from the galactic center and the best $D\left (R\right )$ for each galaxy is obtained by matching the experimental data with the NFDG fits. An alternative method for computing $D(R)$ directly from the galactic mass distributions will be introduced later in this section, but this second method is, at the moment, less effective than the one described above. Full details about the NFDG model can be found in our papers I-V, with the latest version of our computations described in paper V \cite{Varieschi:2022mid}. 

It should be noted that NFDG does not imply any External Field Effect \cite{Varieschi:2022mid}, so that the solar system gravity is essentially due to the Sun’s gravitational field, with other nearby sources at most producing minor tidal forces. 
At the moment, we haven't fully analyzed in NFDG the role of spherically-symmetric sources, in order to explain why standard gravity ($D=3$) is present at the solar system level, or for any other similar star system. This will be done in a future publication which will also study other spherically-symmetric sources, such as globular clusters or galaxies with spherical symmetry.

However, in our previous papers dealing with axially-symmetric sources we have shown how the dynamics of stars orbiting in a disk galaxy depends on the galactic mass distribution, which might be better described by a fractal structure, and thus by an effective NFDG dimension $D \leq 3$, also depending on position. This is the main NFDG hypothesis, supported by how the galactic rotation curves can be effectively described by NFDG methods \cite{Varieschi:2020dnd,Varieschi:2020hvp,Varieschi:2022mid}. The three situations analyzed in this manuscript are meant to provide additional support for this hypothesis. In the following, we will first apply NFDG to the case of AGC 114905.

The gas-rich ultra-diffuse galaxy AGC 114905 has always been considered an outlier of the baryonic Tully-Fisher relation, but the latest observations and analysis of this galaxy \cite{PinaMancera:2021wpc} have shown that its rotation curve can be explained almost entirely by the baryonic mass distribution alone, hence the possibility of the existence of dark-matter-free galaxies. This was not the first galaxy to be shown as lacking dark matter, but since NFDG\ methods outlined above can be easily applied to this case, we will consider AGC 114905 as our first and main example. 

We note that AGC 114905 posed a challenge for other alternative theories of gravity, such as MOND, as shown in the original paper by  Mancera Pi{\~n}a et al. \cite{PinaMancera:2021wpc}. However, subsequent work by Banik et al. \cite{Banik:2022gqc} has shown that the MOND analysis can be consistent with the experimental data, by invoking a much lower inclination angle for this galaxy than the one assumed in Ref. \cite{PinaMancera:2021wpc}.

\begin{figure}\centering 
\setlength\fboxrule{0in}\setlength\fboxsep{0.1in}\fcolorbox[HTML]{000000}{FFFFFF}{\includegraphics[ width=6.99in, height=5.43in,]{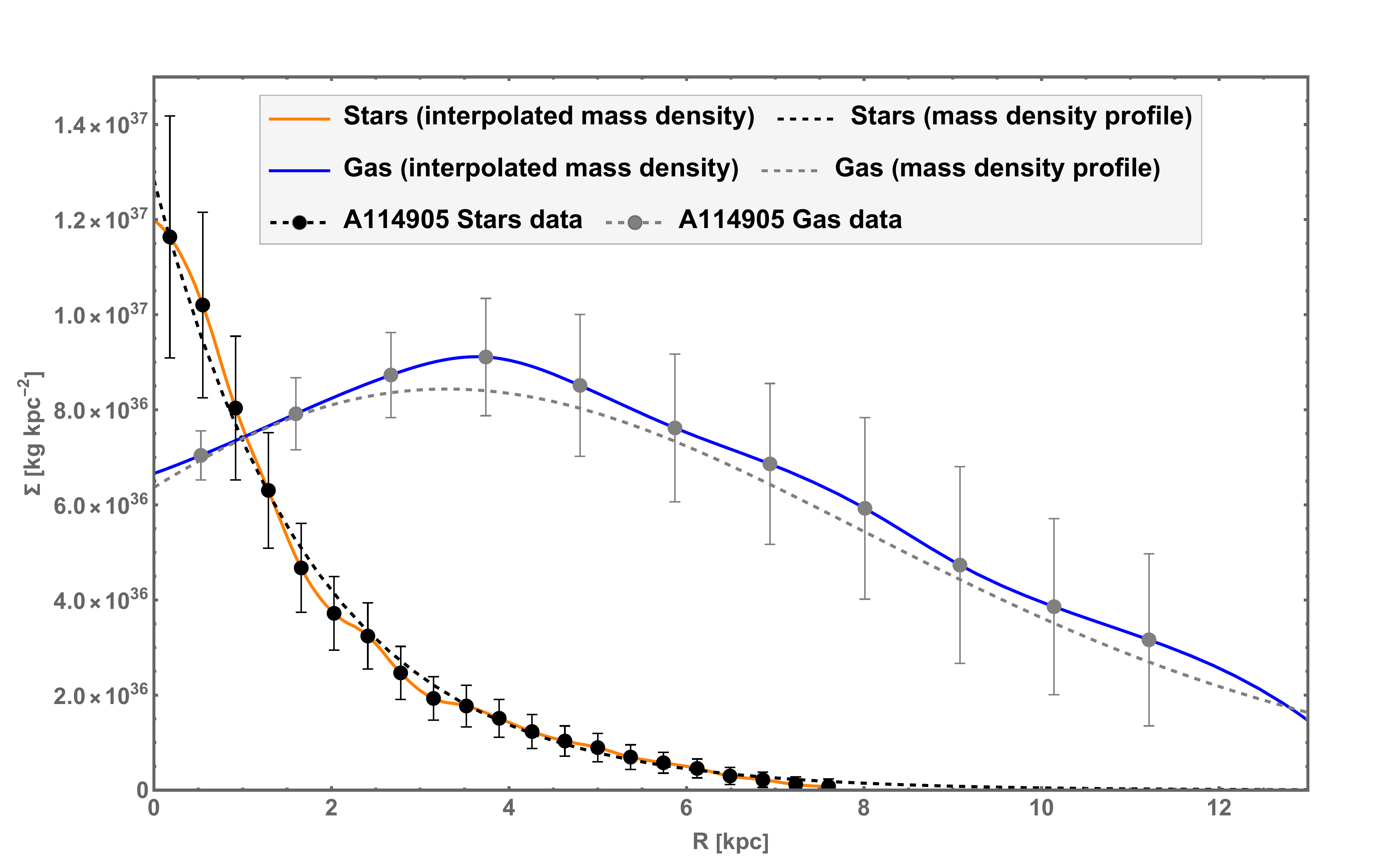}
}
\caption{Surface mass densities for AGC 114905. Data from Mancera Pi{\~n}a et al. \cite{PinaMancera:2021wpc,PinaMancera:2022pri}, for the stellar and gas components, are interpolated with spline functions (solid-orange and solid-blue curves), or with the mass density profiles in Eq. (\ref{eq2.4}) (dashed curves).
}\label{figure:massAGC}\end{figure}

Data and properties of AGC 114905 were obtained from Refs. \cite{PinaMancera:2021wpc,PinaMancera:2022pri}, beginning with the mass modeling for this galaxy. In figure \ref{figure:massAGC}, we show the mass data for the two main baryonic components (Stars data, black circles - Gas data, gray circles) as computed by Mancera Pi{\~n}a et al. \cite{PinaMancera:2021wpc,PinaMancera:2022pri} from the original luminosity data. This figure is equivalent to the right panel of Figure 1 in Ref. \cite{PinaMancera:2021wpc}, although we use different units for the surface mass density $\Sigma $.

These data can then be interpolated with a spline function of appropriate order (solid-orange curve for Stars, solid-blue curve for Gas component), or we can use the density profiles as in Ref. \cite{PinaMancera:2021wpc} (dashed-black curve for Stars, dashed-gray curve for Gas component). These profiles for the stellar disk and the gas component (H{\scriptsize I} plus helium) surface mass densities are defined as:

\begin{gather}\Sigma _{disk}\left (R\right ) =\genfrac{(}{)}{}{}{M_{ \ast }}{2\pi R_{d}^{2}}e^{ -R/R_{d}} \label{eq2.4} \\
\Sigma _{gas}\left (R\right ) =\Sigma _{0 ,gas}e^{ -R/R_{1}}\left (1 +R/R_{2}\right )^{\alpha } , \nonumber \end{gather}where $M_{ \ast } =1.3 \times 10^{8}M_{ \odot }$ and $R_{d} =1.79\ kpc\ $ for the exponential stellar disk profile, while $\Sigma _{0 ,gas} =3.2\ \ M_{ \odot }/pc^{2}$, $R_{1} =1.1\ kpc$, $R_{2} =16.5\ kpc$ and $\alpha  =18$ for the gas profile \cite{PinaMancera:2021wpc}. The first profile in Eq. (\ref{eq2.4}) is a simple exponential function typically used to model stellar disks, while the second profile in the same equation is the best analytical fit for the gas component introduced in Ref. \cite{PinaMancera:2021wpc}.

As in the same Ref. \cite{PinaMancera:2021wpc}, we adopted a $\ensuremath{\operatorname*{sech}}^{2}$ profile along the vertical direction with a constant thickness $z_{d} =0.196\ R_{d}^{0.633} \approx 280\ pc$ \cite{2010ApJ...716..234B} for the stellar disk, and a Gaussian profile with a constant vertical scale height $z_{d} =250\ pc$ for the gas disk. These vertical profiles, together with the radial ones in Eq. (\ref{eq2.4}), were used as the main input for our NFDG mass distributions.\protect\footnote{
The alternative radial profiles given by the direct interpolation of the mass densities (solid-orange and solid-blue curves in Fig. \ref{figure:massAGC}) were also tested, but were found to yield results very similar to those based on the analytical profiles in Eq. (\ref{eq2.4}), so these alternative results will be omitted here.
}

\begin{figure}\centering 
\setlength\fboxrule{0in}\setlength\fboxsep{0.1in}\fcolorbox[HTML]{000000}{FFFFFF}{\includegraphics[ width=6.807in, height=8.50in,]{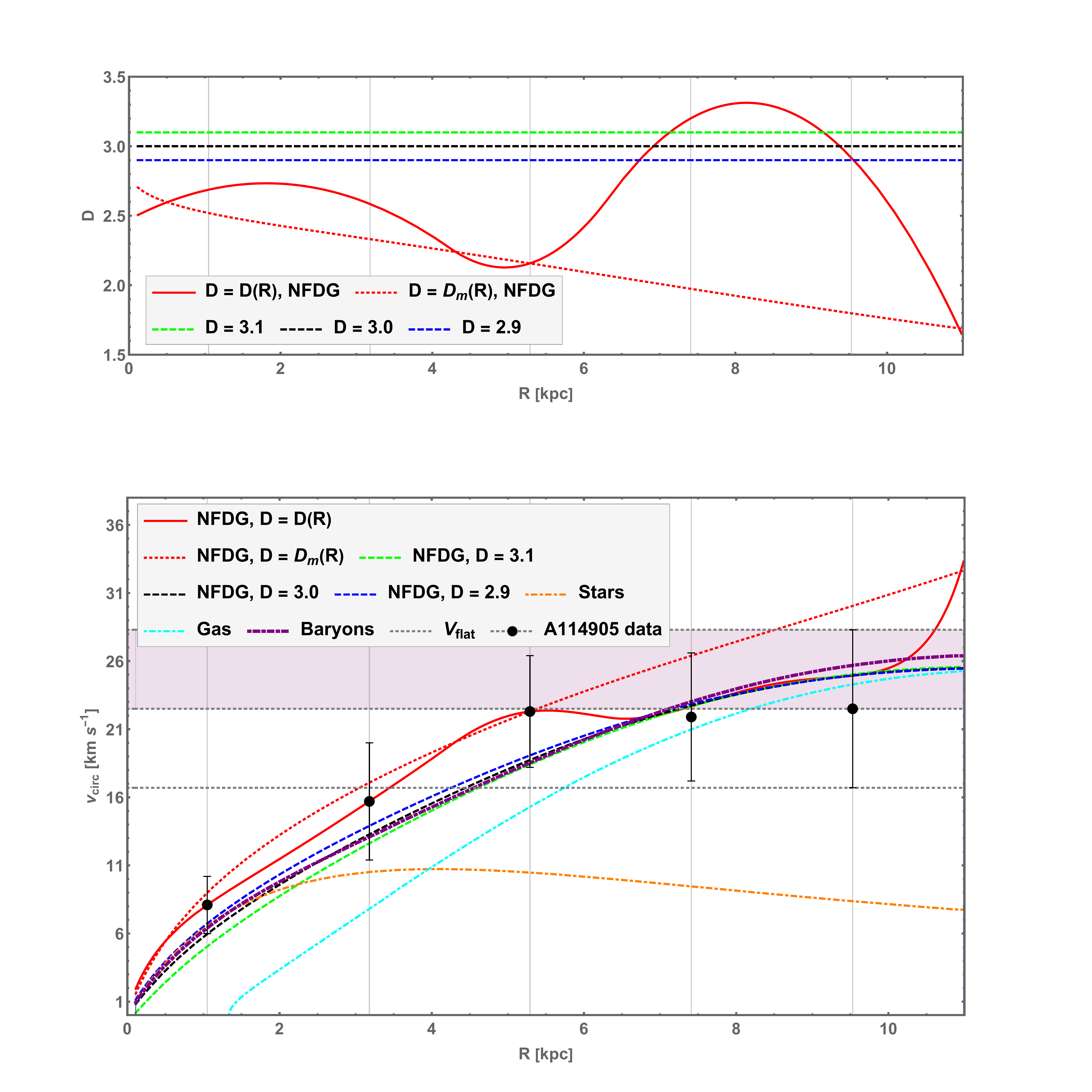}
}
\caption{NFDG results for AGC 114905.
Top panel: NFDG variable dimension $D\left (R\right )$ based directly on the galaxy data (red-solid curve) and NFDG variable dimension $D_m\left (R\right )$ based on the mass dimension of a fractal system (red-dotted curve), are compared with fixed values $D =3.1$, $D =3.0$, and $D =2.9$ (dashed lines). Bottom panel: NFDG rotation curves (circular velocity vs. radial distance) compared to the original data \cite{PinaMancera:2021wpc,PinaMancera:2022pri} (black circles with error bars). The NFDG best fit for the variable dimension $D\left (R\right )$ is shown by the red-solid line, the NFDG fit for $D_m\left (R\right )$ is shown by the red-dotted line, while NFDG fits for fixed values $D =3.1$, $D =3.0$, and $D =2.9$ are shown respectively by the green-dashed, black-dashed, and blue-dashed lines. Also shown: original computation by Mancera Pi{\~n}a et al. \cite{PinaMancera:2021wpc,PinaMancera:2022pri} of the contributions expected from stars (orange dot-dashed line), from gas (cyan dot-dashed line), and total baryonic (magenta, dot-dashed line). The asymptotic flat velocity band (horizontal gray band) is based instead on the data/errors of the last experimental point.
}\label{figure:AGC114905_1}\end{figure}

We then used the latest version of the NFDG\ computation, as detailed in Appendix A of our paper V \cite{Varieschi:2022mid}, in order to obtain the variable dimension function $D\left (R\right )$ which yields the best NFDG fit to the observed circular velocity data for AGC 114905. These experimental data points were obtained from the same sources in Refs. \cite{PinaMancera:2021wpc,PinaMancera:2022pri}, and are shown in Fig. \ref{figure:AGC114905_1} (bottom panel, black circles with error bars), together with the original contributions \cite{PinaMancera:2021wpc,PinaMancera:2022pri} expected from stars (orange, dot-dashed line), gas (cyan, dot-dashed line), and total baryonic (stars plus gas, magenta dot-dashed line). These data and related contributions are the same as shown in Figure 4 of Ref. \cite{PinaMancera:2021wpc}.

The additional curves shown in both panels of Fig. \ref{figure:AGC114905_1} are instead related to our NFDG\ computations. In the top panel we show our best NFDG\  variable dimension function $D\left (R\right )$ as a red-solid line, together with three fixed dimension values $D =3.1 ,\ 3.0 ,\ 2.9$, related to the green, black, and blue dotted lines, respectively. The corresponding NFDG fits to the rotation data are shown in the bottom panel using lines of the same color/type. 

The main NFDG fit is represented by the red-solid line in the bottom panel of the figure. It should be noted that this fit is based on only five experimental data points available for AGC 114905, as opposed to the many experimental points which were available for each of the galaxies studied in our previous papers (based on the SPARC data catalog \cite{Lelli:2016zqa}). Therefore, the NFDG computations of the variable dimension function $D\left (R\right )$ and related rotation velocity (red-solid curves in both panels) were only carried out for the five radial distances of the known experimental data (shown by the vertical gray-thin lines in both panels of Fig. \ref{figure:AGC114905_1}), while interpolation was used in between these radial distances, making our predictions somewhat less effective than results obtained in previous papers.

However, our main NFDG $D\left (R\right )$ fit in the bottom panel can effectively match the experimental data:\  the first three data points are perfectly matched, while for the last two our red-solid curve is close to the data points and well within their respective error bars. The other NFDG $D =3.1 ,\ 3.0 ,\ 2.9$ fits in the bottom panel (green, black, blue dashed curves, respectively) were computed in order to compare these fixed-dimension curves with the experimental data, as well as with the original baryonic curve (magenta, dot-dashed). Our NFDG $D =3.0$ curve (black-dashed) is close to the original baryonic curve (magenta, dot-dashed, from Ref. \cite{PinaMancera:2021wpc}), with minor differences probably due to the different methods used to compute these purely Newtonian components.

Considering instead the NFDG $D =3.1 ,\ \ 2.9$ curves (green and blue, dashed lines) over the radial range, we observe that the spread between these curves becomes progressively less pronounced with increasing radial distances, with all the NFDG\ fits becoming almost equivalent in the radial region between the last two experimental points. Again, this shows some limitations of our NFDG fits for the last two data points, probably due to the lack of more precise experimental data in this outer radial region.

An alternative NFDG computation is also shown by the red-dotted curves in both panels of Fig. \ref{figure:AGC114905_1}. This alternative analysis was based on the discussion in Appendix C of Ref. \cite{Varieschi:2022mid}, where the variable dimension function is considered as a field depending on the radial coordinate $R$ and denoted by $D_m(R)$ in the following. In this approach, $D_m(R)$ is now based on the concept of mass dimension of a fractal system, i.e., the mass dimension of a disk-shaped source is defined as $M\left( R \right) \approx M_{0}{\left( {R}/{R_{0}} \right)}^{D_{m}(R)-1}$, with $M(R)$ being the total baryonic mass within radius $R$, and $R_0$ representing a scale length of the system with related mass $M_0$.

Following these definitions, it was shown in the same Appendix C of Ref. \cite{Varieschi:2022mid} that this new field $D_m(R)$ can be easily computed as:
\begin{equation}D_{m}\left( R \right) \approx 1+\ln\left[ \frac{M\left( R \right)}{M\left( R_0 \right)} \right]/\ln\left[ \frac{R}{R_0} \right] , \label{eq2.5}
\end{equation}
where $M\left( R \right)=2\pi\int_{0}^{R} \Sigma_{tot}\left( R' \right)R'dR'$ and $M_{0}\equiv M\left( R_{0} \right)=2\pi\int_{0}^{R_0} \Sigma_{tot}\left( R' \right)R'dR'$. In this way, the field $D_m(R)$ is predicted directly from the baryonic matter distribution alone (described by the total surface mass density $\Sigma_{tot}\left( R' \right)$) and not based on the observed rotational velocity data points, as was done using instead the previous method for $D=D(R)$ (red-solid curves).

As also discussed in Ref. \cite{Varieschi:2022mid}, the value of $R_0$ (and related mass $M_0$) can be left as a free parameter, or simply chosen by matching one of the experimental data points. We have opted for this simpler solution and just arranged this free parameter to match the third data point in the bottom panel of Fig. \ref{figure:AGC114905_1}. The resulting circular velocity fit (red-dotted curve) in the bottom panel seems to effectively match the first three data points, as well as the fourth one within its error bar, while the fifth experimental point is not well-fitted by this curve. The corresponding red-dotted $D_m(R)$ curve in the top panel, shows a smooth variation of the variable dimension, as opposed to the more pronounced changes of the red-solid $D(R)$ curve, computed with the original NFDG method. We would like to remark that this new method of direct computation of the field $D_m(R)$ is still work in progress and will be improved in future publications.  

Apart from these limitations of our NFDG\ fits in the outer radial region, we conclude that our methods work reasonably well also for the case of AGC 114905, analyzed in this section. We confirm that a NFDG $D =3.0$ curve, i.e. purely baryonic, can fit all the five experimental data within their error bars, as in the original analysis by Mancera Pi{\~n}a et al. \cite{PinaMancera:2021wpc}, since our black-dashed curve is close to the magenta dot-dashed curve in the bottom panel of Fig. \ref{figure:AGC114905_1}. However, an improved fit to the experimental points is obtained by our NFDG variable $D\left (R\right )$  curve, with values for the variable dimension at the radial distances of the experimental points detailed in Table \ref{table:one}.

\begin{table}\centering 
\setlength\fboxrule{0cm}\setlength\fboxsep{0cm}\fcolorbox[HTML]{000000}{FFFFFF}{
\begin{tabulary}{53.97499320000001mm}[c]{|C|C|C|}\hline
$R\left [kpc\right ]$
& $v_{circ}\left [\mbox{km}\ \mbox{s}^{ -1}\right ]$
&
$D\left (R\right )$
\\
\hline
$1.05$
&
$8.1 \pm 2.1$
&
$2.69$
\\
\hline
$3.18$ &
$15.7 \pm 4.3$
& $2.58$
\\
\hline
$5.29$
&
$22.3 \pm 4.1$
&
$2.16$
\\
\hline
$7.41$
&
$21.9 \pm 4.7$
& $3.20$
\\
\hline
$9.53$
&
$22.5 \pm 5.8$
& $2.92$
\\
\hline
\end{tabulary}
}
\caption{
AGC 114905 circular velocity data (with errors) and related radial distances \cite{PinaMancera:2021wpc,PinaMancera:2022pri}, together with the corresponding values of the NFDG\ variable dimension $D\left (R\right )$.
}\label{table:one}\end{table}
From this table and from the top panel of Fig. \ref{figure:AGC114905_1} (red-solid curve), we see that this galaxy is best described in NFDG as having a fractional dimension $D \approx 2.5 -2.7$ at lower radial distances, around the first two experimental points, then decreasing toward $D \approx 2.2$ around the region of the third experimental point, and finally increasing toward $D \approx 2.9 -3.2$ in the outer radial region (last two experimental points). However, the NFDG fits to these last two points are not perfect, as already mentioned above. We conclude that NFDG can effectively model also this galaxy, without any DM contribution, although the quality of our fit is limited by the low number of experimental data points available.

\section{\label{sect:ngc}NFDG and the dynamics of NGC 1052-DF2
}
Another notable case of a galaxy apparently lacking dark matter is the ultra-diffuse galaxy NGC 1052-DF2, recently studied by van Dokkum et al. \cite{vanDokkum:2018vup,vanDokkum:2018ccc,vanDokkum:2022zdd}. By measuring the radial velocities of ten globular-cluster objects within this galaxy, it was shown \cite{vanDokkum:2018ccc} that the intrinsic velocity dispersion of these clusters is $\sigma _{gc} =5.6_{ -3.8}^{ +5.2}\ \mbox{km}\ \mbox{s}^{ -1}$, leading to a $90 \% $ confidence upper limit of $\sigma _{gc} <12.4\ \mbox{km}\ \mbox{s}^{ -1}$, with the likelihood level giving instead $\sigma _{gc} =7.8_{ -2.2}^{ +5.2}\ \mbox{km}\ \mbox{s}^{ -1}$ ($\sigma _{gc} <14.6\ \mbox{km}\ \mbox{s}^{ -1}$). These globular-cluster velocity dispersions are very close to the one expected from stars alone, $\sigma _{stars} =7.0_{ -1.3}^{ +1.6}\ \mbox{km}\ \mbox{s}^{ -1}$, thus implying that this galaxy is very deficient in DM and might be a candidate of a purely baryonic galaxy. 

Later papers \cite{Wasserman:2018scp,Emsellem:2019,Danieli:2019zyi,vanDokkum:2022zdd} improved on the original analysis of NGC 1052-DF2, while confirming the lack of dark matter in this galaxy. In particular, in Ref. \cite{Emsellem:2019} the globular-cluster velocity dispersion was updated as $\sigma _{gc} =10.6_{ -2.3}^{ +3.9}\ \mbox{km}\ \mbox{s}^{ -1}$, in Ref. \cite{Danieli:2019zyi} the stellar velocity dispersion was revised as $\sigma _{stars} =8.5_{ -3.1}^{ +2.3}\ \mbox{km}\ \mbox{s}^{ -1}$, and in Ref. \cite{vanDokkum:2022zdd} the ultra-diffuse galaxy DF2 was seen as part of a group of similar dark-matter free galaxies, probably originated from a bullet-dwarf collision.

It should be noted that the original analysis \cite{vanDokkum:2018vup} of this UDG seemed to be inconsistent with MOND\ predictions. However, the correct MOND predictions should include not only the internal field of the galaxy, but also the external field due to the proximity to the giant parent galaxy NGC 1052. Including this external field effect into the MOND analysis, it was shown \cite{Famaey:2018yif} that MOND\ predictions ($\sigma _{MOND} =13.4_{ -3.7}^{ +4.8}\ \mbox{km}\ \mbox{s}^{ -1}$) are in agreement with the observed velocity dispersions reported above.\protect\footnote{
	For a more comprehensive analysis of NGC 1052-DF2 within the context of four popular alternative theories of gravity (MOND, CG, MOG, and Verlinde's Emergent Gravity) see Ref. \cite{Islam:2019irh}.
}

In NFDG, the EFE is not present \cite{Varieschi:2022mid}, but the non-integer variable dimension $D$ postulated by our model will affect the theoretical computation of the velocity dispersion of NGC 1052-DF2. In this paper we limit our analysis to a simple argument based on the NFDG\ version of the virial theorem, while we will leave to future work a more comprehensive analysis of this topic.

In mechanics, the virial theorem \cite{Clausius:1870} links the average over time of the total kinetic energy $\left \langle T\right \rangle $ of a stable system of particles with the average over time of the total potential energy of the system $\left \langle U\right \rangle $. If the potential energy of a two-particle system is of the form $U\left (r\right ) =\alpha r^{n}$, i.e., proportional to some power $n$ of the inter-particle distance $r$, it is easy to show that the virial theorem implies $2\left \langle T\right \rangle  =n\left \langle U\right \rangle $. For the case of standard inverse-square-distance forces, such as Newtonian gravity, we have $n = -1$ and $2\left \langle T\right \rangle  +\left \langle U\right \rangle  =0$, but in NFDG we have instead $n = -\left (D -2\right )$, in view of Eq. (\ref{eq2.1}), and the NFDG virial theorem becomes:

\begin{equation}2\left \langle T\right \rangle  +\left (D -2\right )\left \langle U\right \rangle  =0. \label{eq3.1}
\end{equation}

A simple application of the virial theorem to a system of $N$ astrophysical objects, each of mass $m^{ \prime }$, distributed uniformly inside a sphere of radius $R$ and total mass $M =Nm^{ \prime }$, yields $\left \langle T\right \rangle  =\frac{1}{2}Nm^{ \prime }\left \langle v^{2}\right \rangle  =\frac{1}{2}M\left \langle v^{2}\right \rangle $ and $\left \langle U\right \rangle  = -\frac{3}{5}\frac{GM^{2}}{R}$.  Using the standard ($D =3$) form of the virial theorem and assuming $\left \langle v^{2}\right \rangle  \approx 3\sigma ^{2}$, the ``Newtonian'' velocity dispersion $\sigma _{Newt}$ can be estimated from the total mass $M$ within the radius $R$ as:

\begin{equation}\sigma _{Newt} \approx \sqrt{\frac{GM}{5R}} . \label{eq3.2}
\end{equation}

Using instead the NFDG gravitational potential in Eq. (\ref{eq2.1}) for a constant dimension $D \neq 2$ and with a rescaled radial coordinate instead of the dimensionless one, i.e., $r \rightarrow r/l_{0}$ ($l_{0}$ is the NFDG scale length), it is straightforward to recompute the time-averaged potential energy of the system as $\left \langle U\right \rangle  = -\frac{2D\ \pi ^{1 -D/2}\Gamma \left (D/2\right )}{\left (D -2\right )\left (D +2\right )}\frac{GM^{2}}{l_{0}\left (R/l_{0}\right )^{D -2}}$, which can be used in the generalized\ form of the virial theorem in Eq. (\ref{eq3.1}) to obtain the following NFDG velocity dispersion:

\begin{equation}\sigma _{NFDG} \approx \sqrt{\frac{2D\ \pi ^{1 -D/2}\Gamma \left (D/2\right )GM}{3\left (D +2\right )l_{0}\left (R/l_{0}\right )^{D -2}}} , \label{eq3.3}
\end{equation}which reduces to Eq. (\ref{eq3.2}) for $D =3$. Details of the computations leading to Eqs. (\ref{eq3.2})-(\ref{eq3.3}) are discussed in Appendix \ref{sect:appa}.

In order to use Eq. (\ref{eq3.3}), we need to specify the value of the NFDG scale length $l_{0}$, currently undefined. In view of the connections of NFDG with MOND, as described in all our previous papers I-V, we can assume $l_{0} =\sqrt{GM_{ref}/a_{0}}$ where $M_{ref}$ is an appropriate reference mass and $a_{0} =1.2 \times 10^{ -10}\mbox{m}\ \mbox{s}^{ -2}$ is the MOND acceleration constant. In this section we simply assume $M_{ref} =M$, the total mass of the spherical system within radius $R$. 

It should be noted that the choice of the value for the scale length $l_{0}$ did not affect the calculations of the rotation curves of AGC 114905 in Sect. \ref{sect:agc}, as well as other rotation curves computed in previous papers, since this quantity cancels out in the computations. However, in this section and in the following one, we need to use an explicit value for $l_{0}$ based on the connection with the MOND acceleration $a_{0}$, as outlined above. Since our first paper \cite{Varieschi:2020ioh}, we argued heuristically that $a_{0} \approx GM/l_{0}^2$, as this connection allowed NFDG to recover the MOND flat rotational speed $V_{f}=\sqrt[4]{GMa_{0}}$, the baryonic Tully-Fisher relation, and other fundamental MOND results. In addition, similar connections were used in other fractional gravity models, such as the one proposed by Giusti et al. \cite{Giusti:2020rul,Giusti:2020kcv,Benetti:2023nrp}, where the scale length was explicitly set as $l_{0} =(2/\pi)\sqrt{GM/a_{0}}$, which is practically equivalent to the NFDG relation reported above.\protect\footnote{This model by Giusti et al. \cite{Giusti:2020rul,Giusti:2020kcv,Benetti:2023nrp} is based on a different way to extend the Poisson equation, $\Delta \Phi(\textbf{r}) = 4 \pi G \rho(\textbf{r})$, into a fractional Poisson equation, $(-\Delta)^{s} \Phi(\textbf{r}) = - 4 \pi G l_{0}^{2-2s} \rho(\textbf{r})$, where $(-\Delta)^{s}$ is the fractional Laplacian and $s \in [1,3/2]$ is the fractional index. The Newtonian case is recovered for $s =1$. This fractional Laplacian is defined in terms of fractional operators, while our NFDG is based on standard operators acting on a metric space of fractional dimension $D$ \cite{Varieschi:2020ioh,Varieschi:2021rzk}.}  Following these considerations, we feel confident that our NFDG relation for $l_{0}$ can be used in our current analysis, at least as a first approximation for our study of globular clusters in this section and of the Bullet Cluster in Sect. \ref{sect:bc}.

Although Eqs. (\ref{eq3.2})-(\ref{eq3.3}) are simplified estimates of the velocity dispersion of a system and should be employed just to determine the order of magnitude of these quantities, we used them in the context of  NGC 1052-DF2 to show how $\sigma $ might depend on the effective fractal dimension $D$ of this galaxy. For example, we considered the original data in Ref. \cite{vanDokkum:2018ccc} assuming that the Newtonian velocity dispersion in Eq. (\ref{eq3.2}) corresponds to the one due to  stars alone, i.e., $\sigma _{Newt} \approx \sigma _{stars} =7.0_{ -1.3}^{ +1.6}\ \mbox{km}\ \mbox{s}^{ -1}$ and also used the estimated total stellar mass as the total mass $M =M_{stars} =2.0_{ -0.7}^{ +1.0} \times 10^{8}M_{ \odot }$. 

For each of these values of $\sigma _{Newt}$ and $M$ (central value and errors) we derived the corresponding radius $R$ from Eq. (\ref{eq3.2}), which was then used in the NFDG\ velocity dispersion in Eq. (\ref{eq3.3}), together with the appropriate scale length $l_{0}$. This last equation was then solved numerically for the variable dimension $D$, assuming that the NFDG velocity dispersion is now set equal to the globular-cluster velocity dispersion, i.e., $\sigma _{NFDG} \approx \sigma _{gc} =7.8_{ -2.2}^{ +5.2}\ \mbox{km}\ \mbox{s}^{ -1}$ \cite{vanDokkum:2018ccc}.

The result of this computation is:\protect\footnote{
It should be noted that the positive/negative error value for $\sigma _{gc}$ corresponds to the negative/positive error value for the dimension $D$, shown in Eq. (\ref{eq3.4}).
}

\begin{equation}D =2.91_{ +0.10}^{ -0.28} , \label{eq3.4}
\end{equation}showing that the increase in the globular-cluster velocity dispersion, compared to the purely baryonic stellar estimate, can be due to an effective decrease of the fractional dimension below the standard $D =3$ value. However, our NFDG estimate in Eq. (\ref{eq3.4}) is still very close to standard $D \approx 3$, indicating that this galaxy has very little mass discrepancy, and thus confirming the results in the literature.

\begin{figure}\centering 
\setlength\fboxrule{0in}\setlength\fboxsep{0.1in}\fcolorbox[HTML]{000000}{FFFFFF}{\includegraphics[ width=6.99in, height=5.43in,]{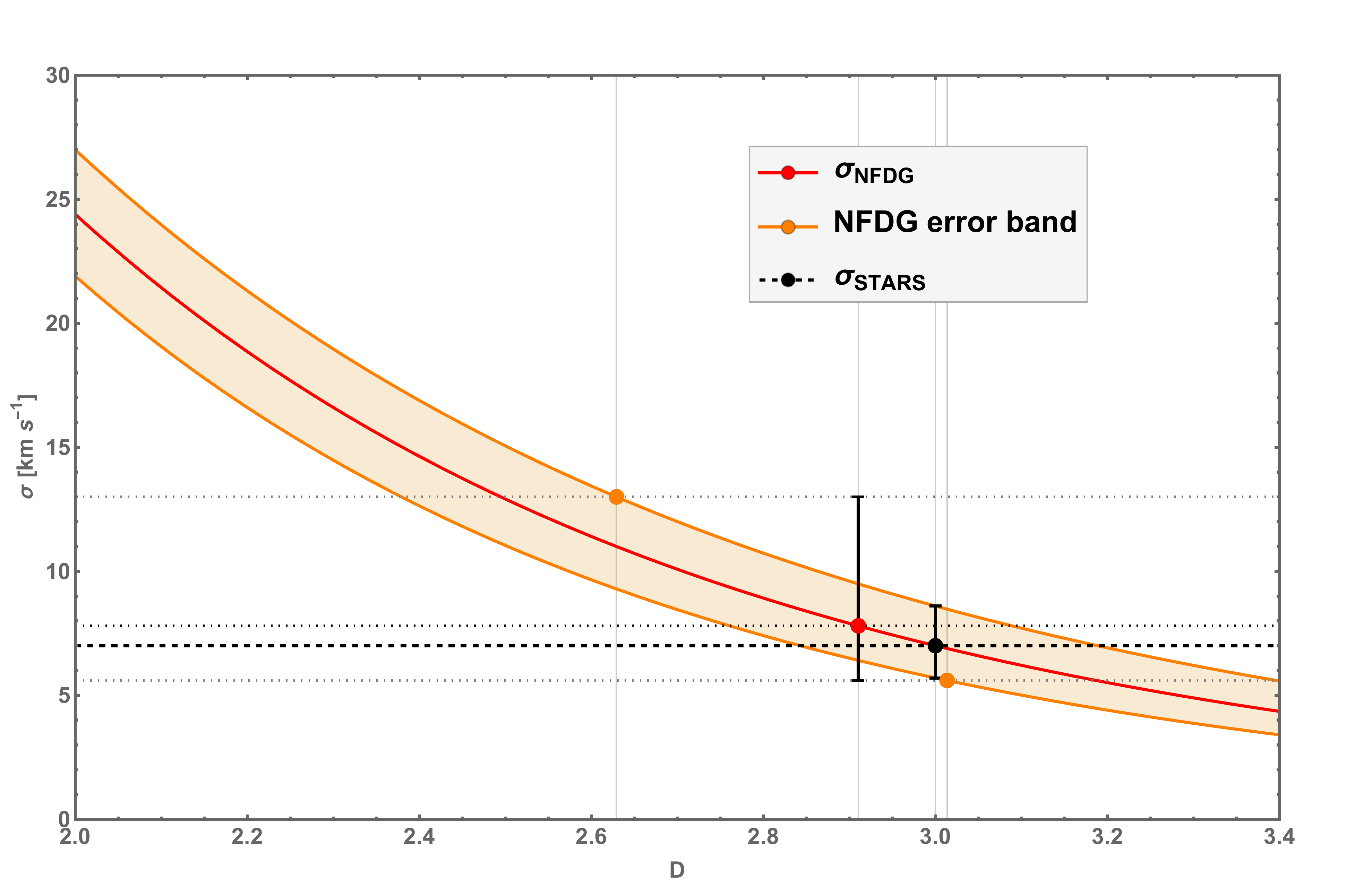}
}
\caption{NFDG velocity dispersion as a function of the variable dimension, following Eq. (\ref{eq3.3}) (red curve and related orange error band). The black-circle data point and related errors are assumed to correspond the stars velocity dispersion for $D =3$, while the red-circle data point and orange-circle error points are assumed to be related to the globular-cluster velocity dispersion with a variable dimension $D$.
}\label{figure:virial}\end{figure}

Figure \ref{figure:virial} provides a more graphical illustration of our argument. In this figure, we show the dependence of the NFDG velocity dispersion $\sigma _{NFDG}$ as a function of the variable dimension $D$, following Eq. (\ref{eq3.3}) and the procedure described above. In particular, the stellar velocity dispersion $\sigma _{stars} =7.0_{ -1.3}^{ +1.6}\ \mbox{km}\ \mbox{s}^{ -1}$ is shown by the black data point corresponding to standard $D =3$ spatial dimension and by the horizontal black-dashed line, while the globular-cluster velocity dispersion $\sigma _{gc} =7.8_{ -2.2}^{ +5.2}\ \mbox{km}\ \mbox{s}^{ -1}$ is shown by the red/orange data points (central value/error points, respectively, also shown with horizontal gray-dotted lines), corresponding to the values of $D$ reported in Eq. (\ref{eq3.4}): $D =2.91$ (red data point), $D =2.63 ,\ 3.01$ (orange error data points).

The vertical gray gridlines correspond to all these values of the dimension $D$, while the main NFDG\ red-solid curve and the related orange error band clearly show how an increase of the observed globular-cluster velocity dispersion, compared to the purely baryonic (stellar) one, could be due to a lower fractional dimension of the galaxy, instead of being attributed to the presence of dark matter.

In other words, this NFDG\ analysis based on the extension of the virial theorem in Eqs. (\ref{eq3.1})-(\ref{eq3.3}), could be used for other galaxies where the discrepancy between baryonic and observed globular-cluster velocity dispersions is usually explained in terms of a DM component. NFDG might be able to explain these discrepancies in terms of a $D \neq 3$ value of the fractional dimension of the galaxy being studied.

It should be noted again that the NFDG approach to globular-clusters dispersion velocity discussed in this section is only a limited, preliminary study. A more comprehensive analysis should be done in NFDG by following methods based on the spherical Jeans equation (see \cite{Wolf:2009tu} and references therein), or others. We will leave this work to future papers on the subject.

\section{\label{sect:bc}NFDG and the Bullet Cluster
}
The Bullet Cluster (BC) merger (1E0657-06) is considered one of the strongest proofs of the existence of DM \cite{Clowe:2006eq,Lage:2014yxa}. This merger is usually described in terms of the gas components of the two colliding clusters (revealed by the X-ray emissions) which, together with standard luminous matter (stars), represent the baryonic content of the cluster pair. In addition, a strong DM component is revealed by gravitational lensing, pointing toward a direct, empirical evidence of dark matter in this collision \cite{Clowe:2006eq}.

A simplified but effective way of modeling this collision \cite{Moffat:2010gt,Calcagni:2021mmj} can be achieved by considering the two clusters as point masses subject to their reciprocal gravitational attraction.\protect\footnote{
	This simplified method discussed in Ref. \cite{Moffat:2010gt} was then linked to the MOG model. A more detailed analysis of the BC merger in MOG can be found in Ref. \cite{Moffat:2012wn}.
}  Using the updated values in Ref. \cite{Lage:2014yxa}, the main cluster is assumed to have a dynamical mass $M_{m} =2 \times 10^{15}M_{ \odot }$, while the bullet cluster dynamical mass is estimated as $M_{b} =5 \times 10^{14}M_{ \odot }$. These ``dynamical'' masses include both baryonic and DM contributions, while purely baryonic masses can be obtained by using the estimated baryonic-to-total mass ratio, $M_{bar}/M_{total} =0.14$ \cite{Lage:2014yxa}, i.e., $M_{m}^{bar} =2.8 \times 10^{14}M_{ \odot }$ and $M_{b}^{bar} =7.0 \times 10^{13}M_{ \odot }$.

The main objective of this simplified analysis is to obtain the initial infall velocity of the two clusters at a certain separation distance, which was estimated in Ref. \cite{Lage:2014yxa} as $2900\ \mbox{km}\ \mbox{s}^{ -1}$ at a distance of $2.5\ $ \textrm{Mpc}. Treating the system of the two point-mass clusters as isolated and with the clusters at relative rest at infinity, the infall velocity computation follows the standard calculation of the escape velocity, by setting to zero both the total energy and the total momentum of the system:

\begin{gather}\frac{1}{2}\left (M_{m}v_{m}^{2} +M_{b}v_{b}^{2}\right ) -G\frac{M_{m}M_{b}}{r} =0 \label{eq4.1} \\
M_{m}v_{m} +M_{b}v_{b} =0. \nonumber \end{gather}

These equations can be easily solved to obtain the standard, Newtonian, infall velocity $(v_{m} -v_{b})_{Newt}$ at the separation distance $r$:

\begin{equation}\left (v_{m} -v_{b}\right )_{Newt} =\sqrt{\frac{2G}{r}\left (M_{m} +M_{b}\right )} . \label{eq4.2}
\end{equation}Using the total cluster masses reported above (including DM) and the separation distance $r =2.5$ \textrm{Mpc}, one obtains $\left (v_{m} -v_{b}\right )_{Newt} =2933\ \mbox{km}\ \mbox{s}^{ -1}$, in line with the expected results \cite{Lage:2014yxa}, while using the purely baryonic masses for the two clusters, also reported above, one would instead obtain $\left (v_{m} -v_{b}\right )_{Newt} =1097\ \mbox{km}\ \mbox{s}^{ -1}$, clearly showing that baryonic masses alone, combined with standard Newtonian gravity, are unable to effectively describe the cluster merger.

However, the same argument can now be used in our NFDG model, by considering the appropriate extension of the gravitational potential energy and using only baryonic masses, since NFDG does not include DM\ components:

\begin{equation}U_{NFDG} = -\frac{2\pi ^{1 -D/2}\Gamma \left (D/2\right )GM_{m}^{bar}M_{b}^{bar}}{\left (D -2\right )l_{0}r^{D -2}} , \label{eq4.3}
\end{equation}which follows from the NFDG gravitational potential for $D \neq 2$ in Eq. (\ref{eq2.1}). Using this NFDG potential energy instead of the Newtonian one in Eq. (\ref{eq4.1}) and solving for the NFDG equivalent infall velocity, we obtain:

\begin{equation}\left (v_{m} -v_{b}\right )_{NFDG} =\sqrt{\frac{4\pi ^{1 -D/2}\Gamma \left (D/2\right )G}{\left (D -2\right )l_{0}\left (r/l_{0}\right )^{D -2}}\left (M_{m}^{bar} +M_{b}^{bar}\right )} . \label{eq4.4}
\end{equation}
In this last equation, we have introduced the rescaled radial coordinate instead of the dimensionless one, i.e., $r \rightarrow r/l_{0}$, with $l_{0}$ representing as usual the NFDG scale length. It is easy to check that Eq. (\ref{eq4.4}) reduces to Eq. (\ref{eq4.2}) for $D =3$. 

We can now assume that the infall velocity is the same as the one obtained with Eq. (\ref{eq4.2}), using the total dynamical masses (including DM) of the two clusters, i.e., set $\left (v_{m} -v_{b}\right )_{NFDG} =2933\ \mbox{km}\ \mbox{s}^{ -1}$ and solve Eq. (\ref{eq4.4}) for the value of the fractional dimension $D$ which would yield the above infall velocity at the same separation distance $r =2.5$ \textrm{Mpc}. Obtaining a reasonable value for $D$ (e.g., $1 \leq D \leq 3$) would show that NFDG can explain the BC dynamics as a fractional-dimension effect, at least within the limits of this simplified model.

In order to solve Eq. (\ref{eq4.4}) numerically, we need again to specify the value of the NFDG scale length $l_{0}$, currently undefined. As it was done in the previous section, we set $l_{0} =\sqrt{GM_{ref}/a_{0}}$, where $M_{ref}$ is an appropriate reference baryonic mass and $a_{0} =1.2 \times 10^{ -10}\mbox{m}\ \mbox{s}^{ -1}$ is the MOND acceleration constant. The reference mass, although arbitrary, should be related to the baryonic masses of the two clusters, with some possible choices being the total baryonic mass $M_{ref} =M_{m}^{bar} +M_{b}^{bar} =3.5 \times 10^{14}M_{ \odot }$, the individual baryonic masses of each cluster ($M_{ref} =M_{m}^{bar} =2.8 \times 10^{14}M_{ \odot }$, or $M_{ref} =M_{b}^{bar} =7.0 \times 10^{13}M_{ \odot }$), the reduced mass of the cluster system ($M_{ref} =M_{m}^{bar}M_{b}^{bar}/\left (M_{m}^{bar} +M_{b}^{bar}\right ) =5.6 \times 10^{13}M_{ \odot }$), etc.

\begin{table}\centering 
\setlength\fboxrule{0cm}\setlength\fboxsep{0cm}\fcolorbox[HTML]{000000}{FFFFFF}{
\begin{tabulary}{53.97499320000001mm}[c]{|C|C|}\hline
$M_{ref}$
& $D$
\\
\hline
$M_{m}^{bar} +M_{b}^{bar}$
&
$2.43$
\\
\hline
$M_{m}^{bar}$
&
$2.45$
\\
\hline
$M_{b}^{bar}$
&
$2.53$
\\
\hline
$\frac{M_{m}^{bar}M_{b}^{bar}}{M_{m}^{bar} +M_{b}^{bar}}$
&
$2.54$
\\
\hline
\end{tabulary}
}
\caption{
NFDG\ fractional dimension $D$ for the Bullet Cluster system for different choices of the reference mass $M_{ref}$. }\label{table:two}\end{table}

Table \ref{table:two} shows the values of the fractional dimension $D$ which would yield in NFDG the infall velocity of $\left (v_{m} -v_{b}\right )_{NFDG} =2933\ \mbox{km}\ \mbox{s}^{ -1}$ for different choices of the reference mass $M_{ref}$, as outlined above. We can see that all these values for $D$ are within a reasonable range of $D \simeq 2.4 -2.5$, similar to values obtained in previous NFDG studies of the dynamics of individual galaxies. Although our analysis of the BC is based on a simplified model of this merger, we conclude that NFDG is potentially able to describe the nature of this phenomenon without using any DM\ component.

\section{\label{sect:conclusion}Conclusions}
In this work, we expanded our NFDG model to include galaxies with little or no dark matter (AGC 114905 and NGC 1052-DF2), as well as the case of the Bullet Cluster merger (1E0657-06), considered one of the strongest proofs for the existence of dark matter. We used different NFDG\ methods for the three cases: our standard NFDG procedures, employed in previous studies, were applied to AGC 114905 and produced an effective fit to the rotational velocity data points without any DM.

For NGC 1052-DF2, we extended the virial theorem to cases with dimension $D <3$, and we were able to show that an increase in the observed velocity dispersion can be the result of a lower overall fractional dimension of the galactic structure. However, for both galaxies our methods yielded $D \lesssim 3$, indicating that Newtonian dynamics can model these galaxies reasonably well and any residual discrepancy can be explained by a decrease of the dimension $D$ below the standard $D=3$ value. In other words, NFDG provides a simple explanation of the almost Newtonian behavior of galaxies such as AGC 114905 and NGC 1052-DF2, in terms of a slight change of the fractional dimension from standard $D=3$, without invoking the EFE and a possible violation of the strong equivalence principle, as it was done by MOND theories when analyzing the same two galaxies.

Finally, a simplified model of the Bullet Cluster merger (1E0657-06) was used to show that the observed infall velocity could be due to a fractional dimension $D \simeq 2.4 -2.5$ of the system, without requiring any DM\  presence. Although the NFDG methods employed in the last two cases (NGC 1052-DF2 and 1E0657-06) will need to be improved and extended, we conclude that our model has shown once again its effectiveness in describing all these astrophysical structures without resorting to the dark matter paradigm.

Therefore, these results constitute another step towards proving that dark matter might not be an exotic component of the Universe, but rather a problem with the standard gravitational theories. However, future NFDG studies are still needed to show that this model can yield other predictions besides galactic rotation curves. Future directions will possibly include a more detailed study of galaxy clusters, gravitational lensing, and an improved analysis of the relativistic version of NFDG \cite{Varieschi:2021rzk}, for possible applications to astrophysics and cosmology.

\begin{acknowledgments}This work was supported by the Seaver College of Science and Engineering, Loyola Marymount University - Los Angeles. The author wishes to
thank Dr. Mancera Pi{\~n}a for sharing AGC 114905 galactic data files and other useful information, and also acknowledge the anonymous reviewers for useful comments and suggestions.

\end{acknowledgments}

\section*{Data Availability}

The datasets generated during and/or analyzed during the current study are available from the corresponding author on reasonable request.

\appendix

\section{\label{sect:appa}NFDG virial theorem and spherical structures}
In this section, we will detail the computations leading to Eqs. (\ref{eq3.2})-(\ref{eq3.3}) in Sect. \ref{sect:ngc}. Let's consider a system of $N$ astrophysical objects, each of mass $m^{ \prime }$, distributed uniformly inside a sphere of radius $R$ and total mass $M =Nm^{ \prime }$. The time-averaged total kinetic energy is $\left \langle T\right \rangle  =\frac{1}{2}Nm^{ \prime }\left \langle v^{2}\right \rangle  =\frac{1}{2}M\left \langle v^{2}\right \rangle  \approx \frac{3}{2}M\sigma ^{2}$, and is related directly to the velocity dispersion $\sigma $.

The NFDG computation of the time-averaged potential energy $\left \langle U\right \rangle $ follows from $d\left \langle U\right \rangle  =\phi _{NFDG}\left (r\right )dm$ with the NFDG potential from Eq. (\ref{eq2.1}), $\phi _{NFDG}\left (r\right ) = -\frac{2\pi ^{1 -D/2}\Gamma \left (D/2\right )}{\left (D -2\right )}\frac{Gm}{l_{0}\left (r/l_{0}\right )^{D -2}}$.  NFDG \cite{Varieschi:2020ioh,Varieschi:2021rzk} is based on the integral of a spherically symmetric function $f =f\left (r\right )$ over a $D$-dimensional metric space $\chi $ \cite{1987JPhA...20.3861S}, $\int \nolimits_{\chi }fd\mu _{H} =\frac{2\pi ^{D/2}}{\Gamma \left (D/2\right )}\int \nolimits_{0}^{\infty }f\left (r\right )r^{D -1}dr$, where $\mu _{H}$ denotes an appropriate Hausdorff measure over the space and the fractional dimension $D$ is assumed to be constant. In view of this integral, the total mass $m\left (r\right )$, within a sphere of radius $r$, is:

\begin{equation}m\left (r\right ) =\frac{2\pi ^{D/2}}{\Gamma \left (D/2\right )}\int \nolimits_{0}^{r}\rho \left (r^{ \prime }\right )l_{0}^{3}\left (r^{ \prime }/l_{0}\right )^{D -1}d\left (r^{ \prime }/l_{0}\right ) , \label{eq5.1}
\end{equation}where the NFDG scale length $l_{0}$ has been added for dimensional correctness, and the standard formula $m\left (r\right ) =4\pi \int \nolimits_{0}^{r}\rho \left (r^{ \prime }\right )r^{ \prime 2}dr^{ \prime }$ is recovered for $D =3$.

If the total mass $M$ is uniformly distributed inside the sphere of radius $R$, i.e., for a constant volume density $\rho $, from the previous equation we have $M =m\left (R\right ) =\frac{2\pi ^{D/2}}{\Gamma \left (D/2\right )}\rho l_{0}^{3 -D}\frac{R^{D}}{D}$, and therefore the constant density $\rho $ can be expressed as $\rho  =\frac{\Gamma \left (D/2\right )}{2\pi ^{D/2}}\frac{DM}{l_{0}^{3 -D}R^{D}}$ ($\rho  =\frac{3M}{4\pi R^{3}}$ for $D =3$). From these equations, it also follows that $m\left (r\right ) =M\genfrac{(}{)}{}{}{r}{R}^{D}$, i.e., the mass within a sphere of radius $r$ obviously scales as $r^{D}$ in NFDG.

Combining all these formulas into the expression for $d\left \langle U\right \rangle $ yields:

\begin{equation}d\left \langle U\right \rangle  =\phi _{NFDG}\left (r\right )dm = -\frac{2\pi ^{1 -D/2}\Gamma \left (D/2\right )}{\left (D -2\right )}\frac{DGM^{2}}{R^{2D}l_{0}^{3 -D}}r^{D +1}dr , \label{eq5.2}
\end{equation}which becomes $d\left \langle U\right \rangle  = -\frac{3GM^{2}}{R^{6}}r^{4}dr$ for $D =3$. Integrating Eq. (\ref{eq5.2}) over the whole sphere, we obtain the time-averaged potential energy already reported in Sect. \ref{sect:ngc}:

\begin{equation}\left \langle U\right \rangle  =\int \nolimits_{sphere}\phi _{NFDG}\left (r\right )dm = -\frac{2D\ \pi ^{1 -D/2}\Gamma \left (D/2\right )}{\left (D -2\right )\left (D +2\right )}\frac{GM^{2}}{l_{0}\left (R/l_{0}\right )^{D -2}} , \label{eq5.3}
\end{equation}which yields the standard result for $D =3$, $\left \langle U\right \rangle  = -\frac{3GM^{2}}{5R}$.

Finally, using the NFDG virial theorem in Eq. (\ref{eq3.1}) together with $\left \langle T\right \rangle  \approx \frac{3}{2}M\sigma ^{2}$ and $\left \langle U\right \rangle $ from Eq. (\ref{eq5.3}), we obtain the final expression in Eq. (\ref{eq3.3}) for the NFDG velocity dispersion $\sigma _{NFDG}$.

\bibliographystyle{apsrev4-1}
\bibliography{RFDGmainNotes}
\end{document}